\documentstyle[aps,tighten,epsf,floats]{revtex}
\begin{document}
\draft
\title{Perturbative Approach to Non-renormalizable
Theories}
\author{ J. Gegelia${ }^a$\footnote{e-mail address:
gegelia@daria.ph.flinders.edu.au}   and  G.Japaridze${ }^b$ }
\address{${ }^a$ School of Physical Sciences, Flinders University of South 
Australia, \\ Bedford Park, S.A. 5042, Australia. \\ 
${ }^b$ Department of Physics, Center for Theoretical studies of Physical 
systems, \\ Clark 
Atlanta University, Atlanta, GA 30314, U.S.} 
\date{\today}
\maketitle
\begin{abstract}
On the perturbatively non-renormalizable and non-perturbatively finite examples
(delta-function type potential in non-relativistic quantum mechanics and
the mathematical model of the propagator by Redmond and
Uretsky in quantum field theory) we illustrate that one can develop a
perturbative approach for
non-renormalizable theory.  The key idea is the
introduction of finite number of 
additional expansion parameters which allows us to 
eliminate all
infinities from the perturbative expressions. The generated 
perturbative series reproduce the expansions of 
the exact analytical solutions. 
\end{abstract}

\newpage
\section{Introduction}

It is well known that divergences itself do not make the main renormalization
problem in quantum
field theories.
One can remove all divergences from any theory performing subtractions,
but these subtractions lead to the ambiguous finite parts.
One can include these arbitrary terms into the finite number of 
physical
parameters only in renormalizable theories, retaining predictive power of the
theory \cite{3}. 

There exists a challenging
possibility of non-perturbative finiteness of non-renormalizable theory
(We recall the idea about non-perturbative finiteness of quantum gravity 
(see for example \cite{Isham}, \cite{nak})).
If a theory is perturbatively non-renormalizable and non-perturbatively finite
in terms
of bare or renormalised parameters (in the latter case 
ambiguities are hidden in 
the renormalised parameters), then the physical 
quantities do not 
contain arbitrary parts and the theory has predictive power. Let 
us assume that we found out somehow that a
perturbatively non-renormalizable theory is non-perturbatively finite, 
but we do 
not have exact solutions. The question we want to address is whether we can
extract any reliable physical information from perturbative expressions of
physical quantities in this theory.

In \cite{gj-1} a new perturbative approach to non-renormalizable quantum field
theory has been suggested. This method introduces a finite number of
additional expansion parameters and assuming non-perturbative finiteness of the
theory gives unambiguous series with finite coefficients for all physical
quantities. Unfortunately at least at the moment being one can not argue that
this series correctly reproduce the features of exact solutions (if they
exist).

 Due to the absence of exact solutions for physically 
relevant field-theoretical models the quantum mechanical examples with 
delta-function type potentials occur to be useful to investigate 
the above mentioned problem.
 This kind of potential, having zero-range or contact interaction, seems to be 
relevant from the point of view of field theory, considering the 
ultraviolet divergences as a trace of short-distance singularities.
Some examples of  regularization  and renormalization of
delta-function potentials in non-relativistic quantum mechanics have been
considered in \cite{jac,beg,gos,man,daniel}.
 
In the present paper we demonstrate that the perturbative approach to 
non-renormalizable quantum field theories suggested in \cite{gj-1} can lead to
consistent results. We consider two examples, delta-function (with derivatives)
potential 
in non-relativistic quantum mechanics and a mathematical model of 
field-theoretical propagator by
Redmond and Uretsky \cite{redmond} and show that the resulting series {\it
reproduce the expansions of exact analytic expressions}.

\section{$\delta$-function type potential in non-relativistic quantum mechanics}

We start to elucidate the procedure of the perturbative treatment of 
non-renormalizable theories on the example of the quantum mechanical 
problem considered in \cite{daniel}. In this example the amplitude is
non-perturbatively finite in terms of two renormalised coupling constants so it
contains only two arbitrary parameters which are 
fixed from two physical
quantities.  This model is perturbatively non-renormalizable i.e. to remove
divergences one has to include
an infinite number of additional (counter-)terms into the potential. The
standard 
perturbative renormalization technique leads to the conclusion that the 
physical quantities depend on an
{\it infinite } number of arbitrary parameters.
The potential has the following form: 

$$
<x|V|x'>=\left[ C+C_2\left( \nabla^2 +\nabla'^2 \right)\right]\delta \left( x-
x'\right)\delta (x)
$$
with two (yet) unspecified parameters $C$ and $C_2$.  One could object that this
$\delta$-type potential is not mathematically well defined.  Note that we do
not seek much physics in
this potential. For our illustrative purposes we could take {\it as our
definition } of the model the cutoff regularized potential with subsequent
removal of
cutoff. We would find that our perturbative approach reproduces the
results of exact solutions. 

The exact
formal expression for the scattering amplitude in s-channel 
for $E\geq 0$ is  (see \cite{daniel}; we take the particle mass $\mu =1$):

\begin{equation}
T(E)={C+C_2^2I_5+
2EC_2\left( 2-C_2I_3\right)\over \left( C_2I_3-1\right)^2-I(E)\left[
C+C_2^2I_5+
2EC_2\left( 2-C_2I_3\right)\right]}
\label{2}
\end{equation}
The integrals

$$
I(E)=2\int {d^3k\over (2\pi )^3}{1\over 2E^+-k^2}={1\over \pi^2}
P\int_0^\infty dk{k^2\over 2E-k^2}-i{1\over 2\pi}\left( 2
E\right)^{1\over 2},
$$
$$
I_3=-2\int {d^3k\over (2\pi )^3}; \ \ \  \ I_5=-2\int {d^3k\over (2\pi )^3}k^2
$$
($E^+\equiv E+i\epsilon $ and $P$ denotes a principal value prescription)
diverge as a linear, third and fifth power of some cut-off regulator.
 So far, the amplitude requires renormalization.

In  \cite{daniel} the renormalization    is carried out by choosing the 
scattering 
length $a$ and the effective range $r_e$ as the renormalization 
parameters and by fixing $C_0$ and $C_2$ demanding 

$$
{1\over T(E)}=-{1\over 2\pi}\left( -{1\over a}+r_eE+O\left(
E^2\right)-i(2E)^{1\over 2}\right)
$$

Our aim is to 
analyse the possibility of extracting meaningful physical information 
from perturbation 
theory for this example and compare perturbative results with exact ones.

Let us bring up some  results for the exact solution. The 
amplitude T(E) after simple and lengthy calculations may be expressed as 

\begin{equation}
T(E)= {x\left( 1+xI_1+2Exy\right)\over 1+xI_1-2EI_1yx^2-
xW(E)\left( 1+xI_1+2Eyx\right)},
\label{11}
\end{equation}
where 
$$
Re(I(E))=I(0)=I_1=-2\int {d^3k\over (2\pi)^3}{1\over k^2}
$$
$x={2\pi a}$, \ \  $y={r_e/(4\pi )}$ and $W(E)=I(E)-I_1=iImI(E)$.

Let us introduce the quantity
\begin{equation}
\alpha^*\left( \mu^2\right) ={1\over 4yx^3}\left[ T\left( \mu^2\right)
-T\left( \mu^2\right]|_{I_1=0}\right)|_{W=0}=
{\left( \mu^2\right)^2xI_1\over 1+xI_1-2\mu^2I_1yx^2}.
\label{13}
\end{equation}
In (\ref{13}) to extract the part which is independent of $W$ we temporarily
considered $W$ independent from $\mu^2$ and put it equal to 0.

Now, extracting $xI_1$ from (\ref{13}) and substituting into expression
(\ref{11}) we get:

\begin{equation}
T\left( E\right)={N\over D},
\label{14}
\end{equation}
where
\begin{equation}
N=x+2Eyx^2-\alpha\left( \mu^2\right)\left[ 2yx^2\left( E-\mu^2\right)
-4E\mu^2y^2x^3\right]
\label{15}
\end{equation}

$$
D=1-xW(E)+2xy\left( \mu^2-E\right)\alpha\left( \mu^2\right)+
2W(E)yx^2\left( E-\mu^2\right)\alpha\left( \mu^2\right)
$$
\begin{equation}
-4y^2x^3E\mu^2W\left( E\right)\alpha\left( \mu^2\right)-
2yx^2W\left( E\right)E
\label{16}
\end{equation}
and 
\begin{equation}
\alpha(E)=\alpha^*\left( \mu^2\right)/\mu^2.
\label{alpha}
\end{equation}
It is straightforward to check that the substitution of the value of 
$\alpha $ from (\ref{alpha}) and (\ref{13}) into 
(\ref{14})-(\ref{16}) leads to  the expression (\ref{11}) 
for $T(E)$. 

Below we demonstrate that, although the above discussed model is 
perturbatively non-renormalizable, we receive  the finite perturbative 
expression for the 
amplitude by introducing an additional expansion 
parameter $\alpha $  and this result is in agreement with the exact solution.

One can easily solve the Lippman-Schwinger equation perturbatively 
for $T(E)$ in $s$-channel and obtain:
$$
T(E)=C+4EC_2+2CC_2I_3+8ECC_2I+C^2I+C^3I^2+C^4I^3+C^5I^4+C^6I^5+
$$
\begin{equation}
C_2^2I_5+6C_2^2EI_3+4C^2C_2II_3+12C^2C_2EI^2+6C^3C_2I^2I_3+
16C^3C_2EI^3+16C_2^2E^2I+...
\label{05}
\end{equation}
Evidently, the expansion of the exact solution (\ref{2}) in $C$ and $C_2$ 
coincides with 
(\ref{05}).
Designating
\begin{equation}
x={2\pi a}=T|_{E=0}, \ \ \ \ 
y={r_e\over 4\pi}={1\over 2x^2}{dReT(E)\over dE}|_{E=0}+{x\over 4\pi^2}
\label{21}
\end{equation}
we can express  $C$ and $C_2$ iteratively from (\ref{21}) and (\ref{05}) as 
power series of $x$
and $y$:
\begin{equation}
C_2={1\over 2}x^2y-x^3yI_1+{3\over 2}x^4yI_1^2-
2x^5yI_1^3-{3\over 8}x^4y^2I_3+...
\label{03}
\end{equation}
and 
\begin{equation}
C=x-x^2I_1+x^3I_1^2-x^4I_1^3-x^3yI_3+x^5I_1^4+3I_1I_3x^4y-x^6I_1^5-
6x^5yI_1^2I_3-{1\over 4}x^4y^2I_5+...
\label{04}
\end{equation}
The substitution of (\ref{03}) and (\ref{04}) into 
(\ref{05}) leads to the following expression 
for the amplitude:
$$
T(E)=x+x^2W(E)+x^3W(E)^2 +x^4W(E)^3 +x^5W(E)^4 +
x^6W(E)^5 +
4EW(E)x^3y+
$$
\begin{equation}
2Ex^2y+6EW(E)^2x^4y+8EW(E)^3x^5y+
4E^2x^4y^2W(E)+4E^2x^4y^2I_1+...
\label{02}
\end{equation}
Now, in analogy with  (\ref{13}) we define
\begin{equation}
\alpha^* \left( \mu^2\right) ={1\over 4x^3y^2}\left[ T\left( \mu^2\right)
-T\left( \mu^2\right)|_{I_1=0}\right]|_{W=0}=\mu^2xI_1+...
\label{22}
\end{equation}
Expressing $xI_1$ from (\ref{22}) and substituting into (\ref{02}) we
get:
$$
T(E)=x+x^2W(E)+x^3W(E)^2 +x^4W(E)^3 +x^5W(E)^4 +
x^6W(E)^5 +
4EW(E)x^3y+
$$
\begin{equation}
2Ex^2y+6EW(E)^2x^4y+8EW(E)^3x^5y+
4E^2x^4y^2W(E)+4E^2x^3y^2{\alpha\left( \mu^2\right)}+...
\label{23}
\end{equation}
where $\alpha\left( \mu^2\right)=\alpha^*\left( \mu^2\right)/\mu^2$.

It is straightforward to check that the perturbative series (\ref{23}) 
is the expansion of the exact result, given by (\ref{14})-(\ref{16}). A 
tedious 
calculation shows that the same statement is true in the next order. It 
should be clear from the above discussion that the generated perturbative 
series in $x$, $y$ and $\alpha$ reproduces the 
expansion of the exact solution up to any orders.

To demonstrate that our approach can have practical sense let us make some
numerical
estimations. For simplicity let us take $\mu =0$.  (\ref{14}) leads to the
following expression:
\begin{equation}
T\left( E\right)={x+2Eyx^2-2yx^2\alpha(0)E\over 1-xW\left( E\right)-
2xy\alpha(0)E+2x^2y\alpha(0)W\left( E\right)E-2x^2yW\left( E\right)E}
\label{nc1}
\end{equation}
Let us take the numerical values for $x$, $y$, $\alpha$ and $E$ satisfying 
\begin{equation}
xy\alpha(0)E<<|ixW\left( E\right)|<1
\label{nc0}
\end{equation}
Under this conditions  
we can approximate (\ref{nc1}) by:
\begin{equation}
T\left( E\right)={x\over 1-xW\left( E\right)}
\label{nc2}
\end{equation}
Expanding (\ref{nc2}) in  $x$  we are led to the convergent series. On the other
hand the perturbative expression under condition (\ref{nc0}) leads to
\begin{equation}
T(E)=x+x^2W(E)+x^3W(E)^2 +x^4W(E)^3 +x^5W(E)^4 +
x^6W(E)^5 +...
\label{nc3}
\end{equation}
This series coincides with the expansion of (\ref{nc2}) and first few terms give
good approximation to the exact result the next corrections being small.
For example, let us take : $\alpha(0)=1$, $x=1$, $xW=-0.1i$ and $xyE=0.001$,
(\ref{nc1}) gives:
$
T={1/(1+0.1i)}=0.(9900)-0.0(9900)i
$
and (\ref{nc3}) leads to:
$
T=(1-0.01+0.0001+...)-i(0.1-0.001+0.00001+...)
$
These simple numerical analysis are just to demonstrate that there exists a
region of numerical values of the parameters where our perturbative approach is
reliable numerically.

\section{A model for field-theoretical propagator} 

To come closer to the quantum field theory problems let us demonstrate new
perturbative approach  on the mathematical model of the propagator of
non-perturbatively finite and perturbatively non-renormalizable field theory
by Redmond and Uretsky
\cite{redmond}. 

Let us consider the following expression 
\begin{equation}
{1\over p^2+\mu^2-i\varepsilon }+{g^2\over M^2}\int_{m_0^2}^{\infty} {dm^2\over 
\left( p^2+m^2-i\varepsilon \right)\left[\left( 1-
g^2{m^2\over M^2}\right)^2+g^4\right]}
\label{a0}
\end{equation}
as a mathematical model of the propagator of some field. Here $\mu$, $M$ and 
$g$ are some parameters and $p$ is momenta. For the simplicity we
take $p^2>0$.
It is easy to check
that the expansion of (\ref{a0}) in terms of $g^2$ produces divergences
and one has to make an infinite number of subtractions to remove them.
Consequently,
final finite perturbative expression contains an infinite number of
arbitrary 
parameters, in other words one would have to introduce an infinite number of
counter-terms into the Lagrangian.
On the other hand it is clear that (\ref{a0}) gives an unique result and 
it does not
contain any ambiguous parameters. 

Below we illustrate that the information extracted from non-renormalizable
divergent perturbative series reproduces results of exact expression.

We omit the first finite term in (\ref{a0}) and define
\begin{equation}
G\left( p^2\right)={g^2\over M^2}I\left( p^2\right)
\label{gganm}
\end{equation}
where
$$
I\left( p^2\right)=\int_{m_0^2}^{\infty} 
{dx\over \left( p^2+x\right)\left[\left( 1-
g^2{x\over M^2}\right)^2+g^4\right]}
$$
Let us introduce a cut-off and calculate the above integral:
$$
\int_{m_0^2}^{\Lambda^2} {dx\over \left( p^2+x\right)\left[\left( 1-
g^2{x\over M^2}\right)^2+g^4\right]}=
$$
$$
{1\over\Delta}\left\{ 2ln{p^2+\Lambda^2\over p^2+m_0^2}
-ln{g^4+\left( 1-g^2{\Lambda^2\over M^2}\right)^2\over g^4+\left( 1-
g^2{m_0^2\over M^2}\right)^2}+{1+g^2{ p^2\over M^2}\over g^2}\left[
arctan{1-g^2{ m_0^2\over M^2}\over g^2}-
arctan{1-g^2{ \Lambda^2\over M^2}\over g^2}\right]\right\}
$$
\begin{equation}
=I(p^2,\Lambda )
\label{a2}
\end{equation}
where
\begin{equation}
\Delta =1+2{g^2\over M^2}p^2+g^4\left( 1+{p^4\over M^4}\right).
\label{5}
\end{equation}
Taking the limit $\Lambda\to \infty $ in (\ref{a2}) we get
$$
\int_{m_0^2}^{\infty } {dx\over \left( p^2+x\right)\left[\left( 1-
g^2{x\over M^2}\right)^2+g^4\right]}=
$$
\begin{equation}
{1\over\Delta}\left\{ 2ln{ M^2\over p^2+m_0^2}
-ln{g^4\over g^4+\left( 1-
g^2{m_0^2\over M^2}\right)^2}+{1+g^2{ p^2\over M^2}\over g^2}\left[
arctan{1-g^2{ m_0^2\over M^2}\over g^2}+{\pi\over 2}\right]\right\}
\label{intsasruli}
\end{equation}
Note that the regularized expression (\ref{a2})
expanded in terms of $g$ has the following structure:
\begin{equation}
\sum_{i,j,k=0,1...}\left( g^2\right)^i
\left( g^2ln{\Lambda^2\over m_0^2}\right)^j
\left( g^2{\Lambda^2\over M^2}\right)^k
C_{ijk}\left(\Lambda^2, M^2, m_0, p^2 \right)
\label{a01}
\end{equation}
where $C_{ijk}$ are finite in the $\Lambda\to \infty $ limit. 
This structure (\ref{a01}) will play an important role in the following 
development.

In (\ref{a2}) we replace
$$
ln{p^2+\Lambda^2\over p^2+m_0^2}=
ln{\Lambda^2\over m_0^2}+ln{m_0^2\over p^2+m_0^2}+ln{\Lambda^2+p^2\over 
\Lambda^2}\to ln{\Lambda^2\over m_0^2}+ln{m_0^2\over p^2+m_0^2}.
$$

Let us define
$$
G\left( p^2,\Lambda \right)={g^2\over M^2}I\left( p^2,\Lambda \right),
$$
the ``related'' quantities (taking the structure (\ref{a01}) into
account, where 
$g^2ln{\Lambda^2\over m_0^2}$ is replaced by $g_1^2ln{\Lambda^2\over m_0^2}$
and $g^2{\Lambda^2\over M^2}$ by $g_2^2{\Lambda^2\over M^2}$): 
$$
M^2\Delta \left( p^2\right)G^*(p^2,\Lambda )=2g_1^2ln{\Lambda^2\over m_0^2}+
2g^2ln{m_0^2\over p^2+m_0^2}
-ln{g^4+\left( 1-g_2^2{\Lambda^2\over M^2}\right)^2\over g^4+\left( 1-
g^2{m_0^2\over M^2}\right)^2}+
$$
$$
+{1+g^2{ p^2\over M^2}\over g^2}\left(
arctan{1-g^2{ m_0^2\over M^2}\over g^2}-
arctan{1-g_2^2{ \Lambda^2\over M^2}\over g^2}\right)
$$
$$
F\left( \lambda_1^2,\lambda_2^2,\Lambda^2\right)=
G^*\left( \lambda_1^2,\Lambda^2\right)\Delta\left( \lambda_1^2\right)-
G^*\left( \lambda_2^2,\Lambda^2\right)\Delta\left( \lambda_2^2\right)=
$$
\begin{equation}
={g^2\over M^2}\left\{ -2ln{\lambda_1^2+m_0^2\over \lambda_2^2+m_0^2}+
{\lambda_1^2-\lambda_2^2\over M^2}\left[
arctan{1-g^2{ m_0^2\over M^2}\over g^2}-
arctan{1-g_2^2{ \Lambda^2\over M^2}\over g^2}\right]\right\},
\label{a4}
\end{equation}
and
\begin{equation}
\alpha ={M^2\over g^2\left( \lambda_1^2-\lambda_2^2\right)}
\left( {M^2\over g^2}F\left( \lambda_1^2,\lambda_2^2,\Lambda^2\right)+
2ln{\lambda_1^2+m_0^2\over \lambda_2^2+m_0^2}\right)
\label{a5}
\end{equation}
From (\ref{a4}) and (\ref{a5}) 
\begin{equation}
tan\left( g^2\alpha\right)=
{g^2g_2^2{\Lambda^2\over M^2}-g^4{m_0^2\over M^2}\over
g^4+\left( 1-g^2{m_0^2\over M^2}\right)\left( 1-
g_2^2{\Lambda^2\over M^2}\right)}.
\label{a6}
\end{equation}
Extracting $g_1^2ln{\Lambda^2\over m_0^2}$ and $g_2^2{\Lambda^2\over M^2}$
from (\ref{a6}) and
$$
G^*\left( \lambda_1^2,\Lambda^2\right)=
{g^2\over M^2}I\left( \lambda_1^2,\Lambda^2\right)
$$
and substituting into $G^*\left( Q^2,\Lambda^2\right)$ we get:
$$
G^*\left( Q^2,\Lambda^2\right)=
$$
$$
={1\over M^2\Delta \left( Q^2\right)}
\left( \Delta\left( \lambda_1^2\right)M^2G^*\left( \lambda_1^2,\Lambda^2\right)
+2g^2ln{ \lambda_1^2+m_0^2\over Q^2+m_0^2}+{Q^2-\lambda_1^2\over M^2}\alpha g^4
\right)
$$
Taking the limit $\Lambda^2\to\infty$ and substituting $g_1=g$, $g_2=g$ we get:
$$
G\left( Q^2\right)=
$$
\begin{equation}
={1\over M^2\Delta \left( Q^2\right)}
\left( \Delta\left( \lambda_1^2\right)M^2G\left( \lambda_1^2\right)
+2g^2ln{ \lambda_1^2+m_0^2\over Q^2+m_0^2}+{Q^2-\lambda_1^2\over M^2}\alpha g^4
\right).
\label{a8}
\end{equation}
Thus we have expressed the finite quantity $G\left( Q^2\right)$ in 
terms of
other finite quantities $g^2$, $G\left( \lambda_1^2\right)$ and $\alpha$.
It is straightforward to check that by substituting $G\left( \lambda_1^2\right)$
and $\alpha $ into (\ref{a8}) we are led to the correct expression (given by
(\ref{gganm}) and (\ref{intsasruli})) for 
$G\left( Q^2\right)$.

On the other hand, expansion of (\ref{a2}) in terms of $g^2$ gives the 
perturbative
expression for $G\left( p^2,\Lambda^2\right)$:
$$
M^2G\left( p^2,\Lambda^2\right)=2g^2ln{\Lambda^2\over m_0^2}-
2g^2ln{m_0^2+p^2\over m_0^2}+2g^4{\Lambda^2\over M^2}-
2g^4{m_0^2\over M^2}-4g^4{p^2\over M^2}ln{\Lambda^2\over m_0^2}+
$$
$$
+4g^4{p^2\over M^2}ln{m_0^2+p^2\over m_0^2}+{3\over 2}g^6{\Lambda^4\over M^4}-
{3\over 2}g^6{m_0^4\over M^4}-3g^6{p^2\Lambda^2\over M^4}+
3g^6{p^2m_0^2\over M^4}+6g^6{p^4\over M^4}ln{\Lambda^2\over m_0^2}-
$$
\begin{equation}
6g^6{p^4\over M^4}ln{m_0^2+p^2\over m_0^2}-2g^6ln{\Lambda^2\over m_0^2}+
2g^6ln{m_0^2+p^2\over m_0^2}+...
\label{a10}
\end{equation}
where we have replaced $ln{p^2+\Lambda^2\over p^2+m_0^2}\to 
ln{\Lambda^2\over m_0^2}+ln{m_0^2\over p^2+m_0^2}$

Let us introduce ``related'' quantities (taking the structure (\ref{a01}) into
account 
$g^2ln{\Lambda^2\over m_0^2}$ is replaced by $g_1^2ln{\Lambda^2\over m_0^2}$
and $g^2{\Lambda^2\over M^2}$ by $g_2^2{\Lambda^2\over M^2}$ in (\ref{a10})): 
$$
M^2G^*\left( p^2,\Lambda^2\right)=2g_1^2ln{\Lambda^2\over m_0^2}-
2g^2ln{m_0^2+p^2\over m_0^2}+2g^2g_2^2{\Lambda^2\over M^2}-
2g^4{m_0^2\over M^2}-4g^2g_1^2{p^2\over M^2}ln{\Lambda^2\over m_0^2}+
$$
$$
+4g^4{p^2\over M^2}ln{m_0^2+p^2\over m_0^2}+{3\over 2}g_2^4g^2{\Lambda^4\over 
M^4}-
{3\over 2}g^6{m_0^4\over M^4}-3g^4g_2^2{p^2\Lambda^2\over M^4}+
3g^6{p^2m_0^2\over M^4}+6g^4g^2_1{p^4\over M^4}ln{\Lambda^2\over m_0^2}-
$$
$$
-6g^6{p^4\over M^4}ln{m_0^2+p^2\over m_0^2}-
2g^4g^2_1ln{\Lambda^2\over m_0^2}+2g^6ln{m_0^2+p^2\over m_0^2}+...
$$
and
$$
M^2F\left( \lambda_1^2,\lambda_2^2,\Lambda^2\right)=
M^2\left\{G^*\left( \lambda_1^2,\Lambda^2\right)\Delta\left( 
\lambda_1^2\right)-
G^*\left( \lambda_2^2,\Lambda^2\right)\Delta\left( \lambda_2^2\right)\right\}=
$$ 
\begin{equation}
=g_2^2g^4{\Lambda^2\over M^4}\left(\lambda_1^2-\lambda_2^2\right)-
g^6{m_0^2\over M^4}\left(\lambda_1^2-\lambda_2^2\right)-g^2ln{m_0^2+
\lambda_1^2\over m_0^2+\lambda_2^2}+...
\label{a13}
\end{equation}
where $\Delta $ is defined by (\ref{5}).

Expressing $g_1^2ln{\Lambda^2\over m_0^2}$ from 
$M^2G^*\left( \lambda_1^2,\Lambda^2\right)$ we get (The structure (\ref{a01})
guarantees that $g_1^2ln{\Lambda^2\over m_0^2}$ and $g^2_2{\Lambda^2\over M^2}$
can be extracted perturbatively from physical quantities up to arbitrarily large
order of expansion parameters):
$$
g_1^2ln{\Lambda^2\over m_0^2}=
{1\over 2}M^2G^*\left(\lambda_1^2,\Lambda^2\right)+
{1\over 2}g^2ln{m_0^2+\lambda_1^2\over m_0^2}+
{\lambda_1^2\over M^2}g^2G^*\left( \lambda_1^2,\Lambda^2\right)-
g^2g^2_2{\Lambda^2\over M^2}+
$$
$$
+{m_0^2\over M^2}g^4+
{1\over 2}\left( 1+
{\lambda_1^4\over M^4}\right)G^*\left( \lambda_1^2,\Lambda^2\right)g^4+
\left( {3\over 4}{m_0^4\over M^4}+{\lambda_1^4m_0^2\over 2M^4}\right)g^6-
{3\over 4}g^2g_2^4{\Lambda^4\over M^4}-
$$
\begin{equation}
-{\lambda_1^2\over M^2}g^4g_2^2{\Lambda^2\over 2M^2}+...
\label{a12}
\end{equation}
From (\ref{a13})
\begin{equation}
g_2^2{\Lambda^2\over M^2}=g{m_0^2\over M^2}+\alpha+...
\label{a14}
\end{equation}
where
$$
\alpha ={M^2\over g^2\left( \lambda_1^2-\lambda_2^2\right)}
\left( {M^2\over g^2}F\left( \lambda_1^2,\lambda_2^2,\Lambda^2\right)+
ln{\lambda_1^2+m_0^2\over \lambda_2^2+m_0^2}\right)
$$
Substituting (\ref{a12}) and (\ref{a14}) into perturbative expression of
$G^*\left( Q^2,\Lambda^2\right)$, taking $g_1=g_2=g$ in the limit
$\Lambda^2\to\infty$ we get:
$$
G\left( Q^2\right)=G\left( \lambda_1^2\right)+
{2g^2\over M^2}ln{m_0^2+\lambda_1^2\over m_0^2+Q^2}+
2{\lambda_1^2-Q^2\over M^2}g^2G\left( \lambda_1^2\right)+
$$
$$
+{\lambda_1^4-3\lambda_1^2Q^2+4Q^4\over M^4}G\left( \lambda_1^2\right)g^4+
{Q^2-\lambda_1^2\over M^4}g^4\alpha+4g^4{Q^2\over M^4}ln{m_0^2+Q^2\over
m_0^2+\lambda_1^2}+
$$
\begin{equation}
{2g^6\over M^2}ln{m_0^2+Q^2\over m_0^2+
\lambda_1^2}\left( 1-3{Q^4\over M^4}\right)+...
\label{a16}
\end{equation}
It is easy to check that (\ref{a16}) coincides to the expansion of 
(\ref{a8}). So the perturbation 
theory reproduces the results of exact solution.


\section{Conclusions}

The perturbative approach to non-renormalizable 
theories \cite{gj-1} based on introduction of a finite number of additional
expansion parameters
correctly reproduces the exact results for perturbatively non-renormalizable and
non-perturbatively finite quantum mechanical problem. If the model under 
consideration would 
be realistic, one could extract the values of the parameters from the 
observables and compare the predictions of (perturbation) theory with 
experimental data, while the standard perturbative renormalization
technique requires the introduction of {\it infinite} number of additional
parameters and the theory has no predictive power. The same is true for the
mathematical model of field-theoretical propagator by Redmond and Uretsky.

Note that there
exists infinite number of
choices for additional expansion parameters. 
 This flexibility is the realization of the freedom 
in choice of renormalization scheme. Surely, the different normalisation
schemes are not 
equivalent from the point of view of numerical convergence.
Although in general the problem of
numerical convergence of perturbative series
(as well, as conventional series, arising in renormalizable theories) 
remains open, we are optimistic about applications of ideas sketched in
\cite{gj-1} to the problems of non-renormalizable quantum field theories and in
particular to quantum (Einstein's) gravity. 

\medskip
\medskip

{\bf ACKNOWLEDGEMENTS} 

This work was carried out whilst one of the authors (J.G.) was a recipient of 
an Overseas Postgraduate
Research Scholarship and a Flinders University Research Scholarship
holder at Flinders University of South Australia.

This work was supported in part by National Science Foundation under 
Grant No. HRD9450386, Air Force Office of Scientific Research under 
Grant No. F4962-96-1-0211 and Army Research Office under Grant No. 
DAAH04-95-1-0651.



\begin{references}
\bibitem{gj-1}
J. Gegelia, G. Japaridze, N. Kiknadze and K. Turashvili,  hep-th/9507037;
\bibitem{3}
J.C.Collins, {\it Renormalization,} Cambridge University Press, 1984;
\bibitem{Isham}
C. J. Isham, {\em Structural Issues in
Quantum Gravity} Report No. gr-qc/9510063;
\bibitem{nak}
N. Nakanishi and I. Ojima, {\it Covariant Operator Formalism of Gauge
Theories and Quantum Gravity,} World Scientific Lecture notes in
Physics, v.27, World Scientific, 1990;
\bibitem{jac}
R.Jackiw, in "M.A.Beg Memorial Volume", World Scientific, Singapore 
1991;
\bibitem{beg}
M.A.B.Beg and R.C.Furlong, Phys.Rev. D{\bf 31}, a370 (1985);
\bibitem{gos}
P.Gosdzinsky and R.Tarrach, Am. J.Phys. {\bf 59},70 (1991);
\bibitem{man}
C.Manuel and R.Tarrach, Phys.Lett. {\bf B328}, 113 (1994);                   
\bibitem{daniel}
Daniel R.Phillips, Silas R.Beane and Thomas D.Cohen, hep-th/9706070;
\bibitem{w1}
S.Weinberg,Phys.Lett. {\bf B251}, 288 (1990);
\bibitem{w2}
S.Weinberg, Nucl.Phys. {\bf B363}, 3 (19910;
\bibitem{kap}
D.B.Kaplan. M.B.Savage and M.B.Wise, Nucl.Phys. {\bf B478}, 629 (1996);
\bibitem{redmond}
P.J.Redmond and J.L.Uretsky, Phys.Rev.Lett. 1958, {\bf 1}, 147;
\end{references}
\end{document}